%% file: Camera_Ready.tex
\documentclass[conference]{IEEEtran}

\IEEEoverridecommandlockouts

\usepackage{graphicx}
\usepackage{color}
\usepackage{cite}
\usepackage{amsmath}
\usepackage[caption=false]{subfig}
\usepackage{amssymb}
\usepackage{comment}
\usepackage[acronym]{glossaries}
\usepackage[top=.75in, left= 0.625in, right = 0.63in, bottom =1in]{geometry}

\include{my_definitions}
\include{my_acronym}

\allowdisplaybreaks

\begin{document}
\title{Phase Configuration Learning in Wireless Networks with Multiple Reconfigurable Intelligent Surfaces}

\author{\IEEEauthorblockN{George~C.~Alexandropoulos$^1$, Sumudu~Samarakoon$^2$, Mehdi Bennis$^2$, and M\'{e}rouane Debbah$^3$}
$^1$Department of Informatics and Telecommunications, National and Kapodistrian University of Athens, Greece\\
$^2$Centre for Wireless Communications, University of Oulu, Finland\\
$^3$Mathematical and Algorithmic Sciences Lab, Paris Research Center, Huawei Technologies France\\
e-mails: alexandg@di.uoa.gr, \{sumudu.samarakoon, mehdi.bennis\}@oulu.fi, merouane.debbah@huawei.com}


%


\maketitle


\begin{abstract}
Reconfigurable Intelligent Surfaces (RISs) are recently gaining remarkable attention as a low-cost, hardware-efficient, and highly scalable technology capable of offering dynamic control of electro-magnetic wave propagation. Their envisioned dense deployment over various obstacles of the, otherwise passive, wireless communication environment has been considered as a revolutionary means to transform them into network entities with reconfigurable properties, providing increased environmental intelligence for diverse communication objectives. One of the major challenges with RIS-empowered wireless communications is the low-overhead dynamic configuration of multiple RISs, which according to the current hardware designs have very limited computing and storage capabilities. In this paper, we consider a typical communication pair between two nodes that is assisted by a plurality of RISs, and devise low-complexity supervised learning approaches for the RISs' phase configurations. By assuming common tunable phases in groups of each RIS's unit elements, we present multi-layer perceptron Neural Network (NN) architectures that can be trained either with positioning values or the instantaneous channel coefficients. We investigate centralized and individual training of the RISs, as well as their federation, and assess their computational requirements. Our simulation results, including comparisons with the optimal phase configuration scheme, showcase the benefits of adopting individual NNs at RISs for the link budget performance boosting.
\end{abstract}

\begin{IEEEkeywords}
Environmental intelligence, learning, multi-layer perceptron, reconfigurable intelligent surfaces, wave control.
\end{IEEEkeywords}

\IEEEpeerreviewmaketitle

\section{Introduction}
Future wireless communication networks are expected to transform to a unified communication, sensing, and computing platform with embedded artificial intelligence and automation, enabling beyond 5-th Generation (5G) service requirements and diverse vertical applications \cite{Saad_6G_2020,Akyildiz_6G_2020}. To accomplish this overarching goal, advances at various aspects of the network design are necessary, including intelligent network orchestration algorithms, highly reconfigurable and wideband front-ends, and smart wireless connectivity schemes. Reconfigurable Intelligent Surfaces (RISs) \cite{huang2019holographic,DMA_2020} constitute a key wireless hardware technology for the recently conceived concept of Electro-Magnetic (EM) wave propagation control \cite{liaskos2018new,huang2019reconfigurable,Marco2019,WavePropTCCN}, which is envisioned to offer manmade manipulation of the wireless communication environment. This low-cost technology enables easy RIS-based coating of various obstacles and objects of the environment, thus, transforming them into network entities with dynamically reconfigurable properties for wireless communications.

Two of the major challenges facing wireless networks incorporating multiple RISs are the channel acquisition and tracking, as well as the dynamic configuration of the RISs' tunable parameters. The recent literature on these challenges\cite{huang2019reconfigurable,Wu_Zhang_2019} mainly considers RISs with nearly passive unit elements \cite{Kaina_2014}, which are usually metallic elements printed on a substrate. Their EM response can be modified in real time using low-cost and low-power consumption electronics such as PIN diodes, varactors, or transistors. Each of these elements can effectively act locally on the phase of an impinging EM field, taming its reflection or absorption. Only very recently \cite{Taha_2019,hardware2020icassp}, RISs equipped with some active elements for enabling channel estimation at their side have been considered. With the passive RIS designs, channel estimation is quite complex involving the estimation of multiple channels simultaneously at the communication ends, i.e., the Transmitters (TXs) or Receivers (RXs). The proposed approaches, which mainly focus on wireless systems with a single RIS (e.g. \cite{Mishra_2019,8879620,8937491,parafac_SAM2020}), consider dedicated control protocols and predetermined phase configuration patterns, and formulate sparse estimation problems. However, the induced overhead with the control signaling and computation for channel estimation is quite high. For the RIS phase configuration task, the vast majority of the available works assumes perfect channel availability and devises suboptimal solutions for the phase profile design problem at hand. The proposed approaches and algorithms for the single- (e.g, \cite{huang2019reconfigurable,Wu_Zhang_2019}), and very recently, the multi-RIS \cite{MultiRIS1,MultiRIS2} cases are usually centralized, iterative, and computationally demanding. 

Aiming at reducing the overhead of conventional optimization approaches and at capturing potential unmodeled RIS-based features, Neural Networks (NNs) have been lately deployed \cite{Taha_2019,huang2019spawc,liaskos2019spawc,Tasos_DNN_CE_2019,Khan_2019,FL_ICC_2020,VR_RIS_RL_Saad_2020,DRL_ICC_2020} to deal with the design challenges of channel estimation and phase configuration. Considering an indoor simulation scenario with a single RIS, \cite{huang2019spawc} was the first work that designed a deep NN to unveil the mapping between the measured coordinate information at a user location and the phase configuration of the RIS unit elements that maximizes the user’s received signal strength. In \cite{Tasos_DNN_CE_2019}, a twin convolutional NN architecture fed by the received pilot signals was presented that enabled channel estimation, again for the single-RIS case. An NN-based symbol-dependent technique for detecting symbols in signals communicated via a single-RIS was proposed in \cite{Khan_2019}. In \cite{FL_ICC_2020}, the authors adopted Federated Learning (FL) to train individual local NNs using channel measurements at multiple users, which were used for deciding the phase configuration at the single-RIS side. Deep reinforcement learning has been used in \cite{DRL_ICC_2020} to maximize the average energy efficiency in the downlink of a single-RIS-assisted cellular system. Recently \cite{VR_RIS_RL_Saad_2020}, a THz network with multiple RISs and RXs, and a single TX was considered, where deep NNs and  Lyapunov optimization were adopted for the sum-rate maximization scheduling problem.

In this paper, we consider multiple RISs assisting a TX-RX communication pair and propose a supervised learning approach for the phase configurations maximizing the achievable rate, which does not require explicit channel estimation. We present Multi-Layer Perceptron (MLP) NN architectures at each individual RIS or at a central controller, which can be trained either with the position information or the instantaneous channel coefficients. We assess the complexity requirements of all presented NN schemes and compare their performance with the optimal phase configuration based on exhaustive search.

\section{Modeling and Problem Formulation}\label{sec:system_model}
\begin{figure}[!t]
	\centering
	\includegraphics[width=0.74\linewidth]{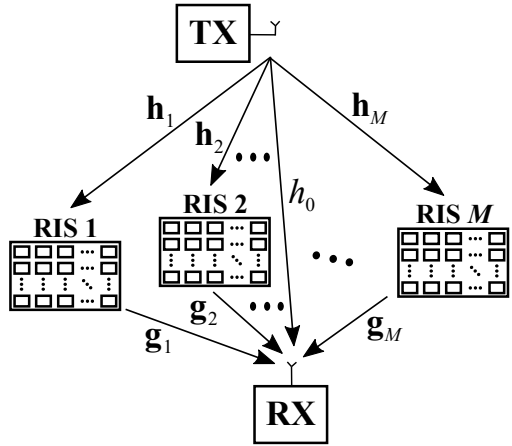}
	\caption{The considered system model for the smart wireless connectivity between a single-antenna Transmitter (TX) and a single-antenna Receiver (RX), which is enabled by $M$ Reconfigurable Intelligent Surfaces (RISs) with similar characteristics, each comprising of $K$ nearly passive unit elements.}
	\label{fig:systemmodel}
\end{figure}

We consider the RIS-empowered wireless communication system illustrated in Fig$.$~\ref{fig:systemmodel}, which consists of a single-antenna TX, a single-antenna RX, and a set $\mathbb{M}$ of $\SURFACE$ similar RISs. Each $m$-th RIS with $m=1,2,\ldots,M$ consists of a two-dimensional rectangular grid $\mathbb{K}$ of $\ELEMENT$ unit elements, which can alter the phase of any impinging EM field \cite{Kaina_2014}. We assume that the direct link between TX and RX and their $M$ two-hop links via the multiple RISs contribute to their wireless communication.  

\subsection{Channel Model}
Let $\mathbf{h}_m\in\mathbb{C}^{K\times 1}$ and $\mathbf{g}_m\in\mathbb{C}^{1\times K}$ represent the wireless channel gains between the $m$-th RIS and TX, and between RX and the $m$-th RIS, respectively, where $\mathbb{C}$ denotes the set of complex numbers. We also use the zero-mean complex Gaussian random variable $h_0$ having variance $L_{d_h}$ to model the Rayleigh faded TX-RX channel gain, where $L_{d_h}$ is the pathloss attenuation depending on these nodes' distance. Similar to \cite{Basar2020}, we assume that all $M$ RISs are located close to RX, and consequently, each $\mathbf{g}_m$ is modeled as a pure Line-Of-Sight (LOS) channel. On the other hand, each $\mathbf{h}_m$ channel is modeled as Ricean faded due to the presence of $S$ scattering objects among TX and RISs, and as such, it is composed of a LOS channel and multiple non-LOS components. In particular, each $K$-element $\mathbf{g}_m$ can be expressed as \cite[eq. (8)]{Basar2020}
\begin{equation}\label{eqn:model_g}
	\mathbf{g}_m \triangleq \sqrt{{\rm G}\left(\theta_{\rm L}\right)L_{d_{\mathbf{g}_m}}}\exp\left(j\eta\right)\mathbf{a}_m\left(\phi_{\rm L},\theta_{\rm L}\right),
\end{equation}
where $\phi_{\rm L}$ and $\theta_{\rm L}$ are the azimuth and elevation Angles Of Arrival (AOAs) of the LOS component with respect to the RIS broadside and $\mathbf{a}_m\left(\phi_{\rm L},\theta_{\rm L}\right)\in\mathbb{C}^{1\times K}$ is the array response vector for the $m$-th rectangular RIS with inter-element spacing $d$. In addition, $\eta$ is uniformly distributed in $[0,2\pi]$ modeling the random phase term induced by the LOS channel, ${\rm G}\left(\theta_{\rm L}\right)$ represents the rotationally symmetric RIS radiation pattern in the $\theta_{\rm L}$ angle, and $L_{d_{\mathbf{g}_m}}$ denotes the pathloss attenuation that depends on the distance $d_{\mathbf{g}_m}$ between RX and the $m$-th RIS. In an analogous way, the Ricean faded $\mathbf{h}_m$ $\forall$$m$ is given by
\begin{equation}\label{eqn:model_h}
	\mathbf{h}_m \triangleq \kappa\sum_{c=1}^C\sum_{r=1}^{R_c}\alpha_{c,r}\Xi_{c,r}\mathbf{a}_m\left(\phi_{c,r},\theta_{c,r}\right)+\beta\mathbf{h}_{m,{\rm L}},
\end{equation}
where $C$ denotes the number of clusters resulting from the $S$ scatterers, with each $c$-th cluster ($c=1,2,\ldots,S$) having $R_c$ sub-rays such that $S=\sum_{c=1}^{C}R_c$. In this expression, we use the notations $\kappa\triangleq S^{-1/2}$ and $\Xi_{c,r}\triangleq\sqrt{{\rm G}\left(\theta_{c,r}\right)L_{c,r}}$ with $\phi_{c,r}$ and $\theta_{c,r}$ being the azimuth and elevation AOAs of the $(c,s)$-th propagation path with respect to the RIS broadside, 
and $\alpha_{c,r}$ represents this path's zero-mean and unit-variance complex Gaussian distributed gain. ${\rm G}\left(\theta_{c,r}\right)$ is the RIS radiation pattern in the $\theta_{c,r}$ angle and $L_{c,r}$ is the pathloss attenuation for the $(c,s)$-th path. Finally, $\beta$ is a Bernoulli random variable characterizing the existence of a LOS link between the $m$-th RIS and RX, and $\mathbf{h}_{m,{\rm L}}\in\mathbb{C}^{K\times 1}$ represents the LOS component of $\mathbf{h}_m$. This vector can be expressed similar to \eqref{eqn:model_g} for the azimuth AOA $\varphi_{\rm L}$ and the elevation AOA $\vartheta_{\rm L}$ of the LOS component with respect to the RIS broadside, and the distance $d_{\mathbf{h}_m}$ between each $m$-th RIS and TX. 

\subsection{Problem Formulation}
The complex-valued baseband received signal at RX via the network of $M$ RISs can be mathematically expressed as 
\begin{equation}\label{system_model}
  y = \left(\sum\limits_{m=1}^M\mathbf{g}_m\mathbf{\Phi}_m\mathbf{h}_{m} + h_0\right)s+n,
\end{equation}
where $\mathbf{\Phi}_m\triangleq\mathrm{diag}\{\boldsymbol{\phi}_m\}\in\mathbb{C}^{K\times K}$ is a diagonal matrix, with $\boldsymbol{\phi}_m\in\mathbb{C}^{K\times 1}$ in the main diagonal, accounting for the effective phase shifts applied by the $K$ unit elements of the $m$-th RIS and $\mathbf{h}_0$ is the channel of direct TX-RX link.
The phase configuration of each $k$-th unit element ($k=1,2,\ldots,K$) of each $m$-th RIS ($m=1,2,\ldots,M$) is modeled as $[\boldsymbol{\phi}_m]_k=e^{j\theta_{m,k}}$ with $j\triangleq\sqrt{-1}$ being the imaginary unit. In this paper, we consider the practical case of finite resolution phase shifting values for the unit elements of each $m$-th RIS, and particularly, each reflection coefficient $[\boldsymbol{\phi}_m]_k$ is obtained as
\begin{equation}\label{phase shifting_model}
[\boldsymbol{\phi}_m]_k \in \mathbb{F} \triangleq \left\{ e^{j2^{1-q}\pi f} \right\}_{f=0}^{2^{q}-1},
\end{equation}
where $\mathbb{F}$ represents each cell's feasible set of reflection coefficients and $q$ is the phase resolution in number of bits. Clearly, the number of phase shifting values per RIS unit element is $2^q$. Finally, the notation $s$ in \eqref{system_model} denotes the complex-valued information symbol of average power $P$ (usually chosen from a discrete constellation set), and $n$ models the zero-mean complex Additive White Gaussian Noise (AWGN) with variance $\sigma^2$. 

Based on \eqref{system_model}'s received signal model, the achievable rate performance per unit bandwidth for the TX-RX communication via the $M$ similar $K$-element RISs is given by
\begin{equation}\label{eqn:datarate}
	\mathcal{R}\left(\{\boldsymbol{\phi}_m\}_{m=1}^M\right) 
	\!=\! 
	\textstyle \log_2 \left( 1 + \frac{P}{\sigma^2}\left|\sum\limits_{m=1}^M\mathbf{g}_m\mathbf{\Phi}_m\mathbf{h}_{m} \!+\! h_0\right|^2\right).
\end{equation}
Hence, the design of the phase configurations for all $M$ RISs that maximizes the rate performance can be obtained from the solution of the following constrained optimization problem:   
\begin{subequations}\label{eqn:maximize_datarate}
\begin{eqnarray}  
	\label{eqn:maximize_datarate_objective}
	\max_{\{\boldsymbol{\phi}_m\}_{m=1}^M} && \mathcal{R}\left(\{\boldsymbol{\phi}_m\}_{m=1}^M\right) \\
	\label{eqn:maximize_datarate_cns}
	\subjectTo && [\boldsymbol{\phi}_m]_k \in \mathbb{F}	\qquad \forall \element\in\mathbb{K}, \forall\surface\in\mathbb{M}.
\textsl{}\end{eqnarray}
\end{subequations}
As a consequence of the constraint \eqref{eqn:maximize_datarate_cns}, even under the perfect knowledge of the channel state information for all involved wireless links (i.e., all $\mathbf{h}_{m}$'s and $\mathbf{g}_{m}$'s, and $h_0$), determining the optimum phase configurations for all $M$ RISs becomes a computationally exhaustive combinatorial problem.

\section{Phase Configuration via Supervised Learning}\label{sec:supervised_learning_solution}
Similar to \cite{huang2019reconfigurable}, we consider RISs equipped with the nearly passive unit elements of \cite{Kaina_2014}, where it is impractical for each individual $m$-th RIS to measure the wireless channels it is involved, i.e., the RIS-TX channel $\mathbf{h}_{m}$ and the RX-RIS channel $\mathbf{g}_{m}$. Inspired by the single-RIS NN approach in \cite{huang2019spawc}, we focus on identifying the underlying relations among the channels and the relative positions of TX, RX, and the $M$ RISs in order to exploit it for designing a low-complexity phase configuration decision maker at each RIS. In mathematical terms, the desired mapping between locations and the optimal phase configuration decisions at each $m$-th RIS can be compactly expressed as:
\begin{equation}\label{eqn:mapping}
	\actionVec_\surface\optimal 
	=  \mapping_{\modelParam_\surface} ( \distanceVecIn_\surface, \distanceVecOut_\surface, \distanceVec ),
\end{equation}
where $\modelParam_\surface$ is a real-valued
vector of tunable parameters for the unknown function $\mapping_{\modelParam_\surface}(\cdot)$, whose size needs to be specified by the learning process. The $3$-Dimensional (3D) positioning vectors $\distanceVecIn_\surface$, $\distanceVecOut_\surface$, and $\distanceVec$ represent the relative positions of TX and RX with respect to the $m$-th RIS, and the relative position of RX with respect to TX, respectively. In this paper, we assume that TX and the $M$ RISs are at fixed positions, hence $\distanceVecIn_\surface$ $\forall$$\surface$ is known, and that $\distanceVec$ can be estimated at each RX new position with conventional localization methods. In addition, $\distanceVecOut_\surface$ $\forall$$\surface$ is assumed to be estimated via recently proposed RIS-based positioning schemes (e.g., \cite{Henk_20}). Alternatively, the RX location can be accurately estimated using RISs equipped with single receive radio-frequency chains \cite{hardware2020icassp}; a detailed method will be described in the extended version of this paper. 

\subsection{Learning via the Positioning Values}
To determine the unknown function in \eqref{eqn:mapping} for each $m$-th RIS, we resort to NN-based function regression and present both a CENtralized (CEN) and a INDividual (IND) supervised learning approaches. For the training referring to the IND approach, we utilize a set $\mathbb{T}$ of labeled training samples $\{\distanceVecIn_{\surface,\tau}, \distanceVecOut_{\surface,\tau},\actionVec_{\surface,\tau}^{\star},\distanceVec_{\tau}\}_{\tau\in\mathbb{T}}$ for each $m$-th RIS and design $\modelParam_\surface$ in order to minimize the following Mean Squared Error (MSE) cost function defined over the training dataset:  
\begin{multline}\label{eqn:lable}
\modelParam_\surface\optimal 
= \argmin\limits_{\modelParam_m} \left(\size{\mathbb{T}}^{-1} 
\sum_{\tau\in\mathbb{T}} \left\| \mapping_{\modelParam_m} \left(\distanceVecIn_{\surface,\tau}, \distanceVecOut_{\surface,\tau},\distanceVec_{\tau}\right) \right.\right. \\
\left.\left.- \actionVec_{\surface,\tau}^{\star} \right\|^2_2 + \lambda \left\|\boldsymbol{w}_m\right\|^2_2 \right).
\end{multline}
In the latter expression, $\size{\mathbb{T}}$ denotes the cardinality of the dataset $\mathbb{T}$ and $\lambda>0$ is a regularization coefficient.
Each $\actionVec_{\surface,\tau}^{\star}$ needed in \eqref{eqn:lable} is obtained from the maximization of $\mathcal{R}\left(\boldsymbol{\phi}_{m,\tau}\right) = \log_2 \left( 1 + P\sigma^{-2}|\mathbf{g}_m\mathbf{\Phi}_m\mathbf{h}_{m}|^2\right)$ subject to the constraint $[\boldsymbol{\phi}_{m,\tau}]_k \in \mathbb{F}$ $\forall$$\element$ RIS unit element. After the NN training phase for computing $\modelParam_\surface\optimal$ from $\mathbb{T}$, each RIS can infer its phase configuration decision for each 3D position, given by the tuple ${\distanceVecIn_\surface}'$, ${\distanceVecOut_\surface}'$, and $\distanceVec'$, as $\actionVec_\surface' = \mapping_{\modelParam_\surface\optimal} ({\distanceVecIn_\surface}', {\distanceVecOut_\surface}',\distanceVec')$.  

For the CEN approach, we consider the following desired mapping, instead of the one in \eqref{eqn:mapping}, to be unveiled from the NN: $\boldsymbol{\varphi}\optimal=\mapping_{\mathbf{w}_0} (\{\distanceVecIn_\surface, \distanceVecOut_\surface\}_{m\in\mathbb{M}}, \distanceVec)$, where the real-valued $\mathbf{w}_0$ denotes the network's tunable parameters and vector $\boldsymbol{\varphi}\optimal\in\mathbb{C}^{KM\times1}$ includes the optimal phase configuration decisions for all $M$ RISs in column concatenation. A labeled dataset $\{\{\distanceVecIn_{\surface,\tau}, \distanceVecOut_{\surface,\tau}\}_{m\in\mathbb{M}},\actionVec_{\tau}^{\star},\distanceVec_{\tau}\}_{\tau\in\mathbb{T}}$ collected from all $M$ RISs is used for the NN training, and $\mathbf{w}_0$ is designed as follows: 
\begin{multline}\label{eqn:sumlable}
	\mathbf{w}_0\optimal =
	 \argmin\limits_{\mathbf{w}_0} \left(
	 \lambda \|\mathbf{w}_0\|^2_2 
	 +
	 \size{\mathbb{T}}^{-1} 
	 \sum_{\tau\in\mathbb{T}} \left\| \boldsymbol{\varphi}_{\tau}^{\star}\right.
	 \right. \\
	 \left. \left.
	 - 
	 \mapping_{\mathbf{w}_0} \left(\{\distanceVecIn_{\surface,\tau}, \distanceVecOut_{\surface,\tau}\}_{m\in\mathbb{M}},\distanceVec_{\tau}\right)
	  \right\|^2_2 \right),
\end{multline}
The $\boldsymbol{\varphi}_{\tau}^{\star}$'s needed in \eqref{eqn:sumlable} can be obtained solving \eqref{eqn:maximize_datarate} using exhaustive search. Similar to the IND approach, when $\modelParam_0$ is computed, the designed NN can be used for inferring the phase configurations of all $M$ RISs for a given RX position.

\begin{figure}[!t]
    \centering
    \includegraphics[width=\columnwidth]{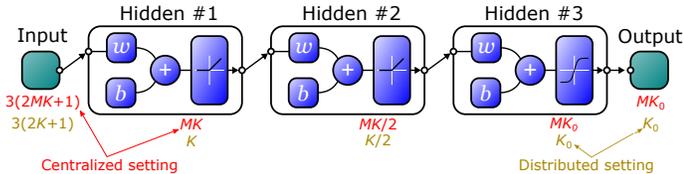}
    \caption{The NN design for both the centralized and the individual RISs approaches including three hidden layers with different numbers of parameters per approach. In the centralized case, all 3D positioning vectors are used as the NN's inputs, which then outputs the phase configurations for all $M$ RISs. The 3D positioning vector referring to the $m$-th RIS is used as the NN's input in the individual RIS case, which outputs the phase configurations for this RIS.}
    \label{fig:nn_setting}
\end{figure}

In Fig$.$~\ref{fig:nn_setting}, we illustrate the proposed MLP architecture for the proposed NN design, considering both the CEN and IND approaches. Starting with the former approach, the input size is $3(2\SURFACE\ELEMENT + 1)$ that depends on the 3D positioning vector with the relative location information for all RISs, the 3D position vector that refers to the relative location between TX and RX, and the included factor $2$ describes the incident and reflection links at all RISs. For the computational tractability of the targeted NN and the RIS phase configuration control architecture, we have assumed that the unit elements in each RIS are divided into $\ELEMENT_0$ groups, such that at each group, the phase configuration is common. For example, supposing that $K/\ELEMENT_0$ is an integer and using \eqref{phase shifting_model}, the number of phase shifting values per $i$-th group of each RIS with $i=1,2,\ldots,K/\ELEMENT_0$ is $2^q$. Using this assumption, the output of the proposed NN in Fig$.$~\ref{fig:nn_setting} for the CEN approach, referring to the decisions for the phase configurations for all $M$ RISs, is of size $\SURFACE\ELEMENT_0$. In addition, the NN model is composed of three fully-connected hidden layers (this was the best NN setup in terms of accuracy versus complexity among the various tested cases) with respective sizes $3\SURFACE\ELEMENT$, $3\SURFACE\ELEMENT/2$, and $\SURFACE\ELEMENT_0$, excluding the bias $b$, and with two Rectified Linear Unit (ReLU) activation layers followed by an activation based on the $\tanh(\cdot)$ function. Note that the CEN approach requires a central entity (could be attached to TX or to one of the RISs) to collect all 3D positioning vectors needed in \eqref{eqn:sumlable} and implement the NN design with the aforedescribed computing and storage capabilities. 

The NN architecture depicted in Fig$.$~\ref{fig:nn_setting} can be also used to realize the proposed IND approach, according to which each individual RIS decides its own phase configuration, independently of the other ones. In contrast to the CEN approach, the IND approach requires that each RIS is equipped with basic computing and storage capabilities to perform supervised learning. In this case, each $m$-th RIS needs to train its local NN with the dataset $\mathbb{T}^\cdot$ of labeled training samples $\{\distanceVecIn_{\surface,\tau}, \distanceVecOut_{\surface,\tau},\actionVec_{\surface,\tau}^{\star},\distanceVec_{\tau}\}_{\tau\in\mathbb{T}^\cdot}$. Hence, the dimensions of each $m$-th NN's input and output are $3(2\ELEMENT+1)$ and $\ELEMENT_0$, respectively, whereas the dimensions of the three hidden layers (without the bias node) are $3\ELEMENT$, $3\ELEMENT/2$, and $\ELEMENT_0$, respectively. These values indicate the storage requirements per individual RIS as well as the required computing capabilities for realizing the proposed MLP-based NN at each RIS side.

\subsection{Learning via the Channel Coefficients}
For comparison purposes, we consider the deployment of the proposed NN architecture with the CEN approach for the case where the network's inputs are the channel coefficients, instead of the positioning values. In particular, we feed the inputs of the network with the phase and magnitude of each channel coefficient corresponding to each RIS unit element and the direct TX-RX link. This case is expected to result in an upper bound performance for the position-based learning, since it requires more demanding and more informative (for the considered system) inputs for the NN. Recall, however, that acquiring channel estimation with passive RISs is a very challenging task, as previously discussed in Section~II. 
In this channel-based learning method, the training dataset $\mathbb{T}^\prime$ is $\{\{\mathbf{h}_{\surface,\tau}, \mathbf{g}_{\surface,\tau}\}_{m\in\mathbb{M}},\actionVec_{\tau}^{\star},h_{0,\tau}\}_{\tau\in\mathbb{T}^\prime}$, and
similar to \eqref{eqn:sumlable}, the considered MSE loss function for the NN design is given by:
\begin{multline}\label{eqn:sumlable_channel}
	\mathbf{w}_0\optimal =
	 \argmin\limits_{\mathbf{w}_0} \left(
	 \lambda \|\mathbf{w}_0\|^2_2 
	 +
	 \size{\mathbb{T}^\prime}^{-1} 
	 \sum_{\tau\in\mathbb{T}} \left\| \boldsymbol{\varphi}_{\tau}^{\star}\right.
	 \right. \\
	 \left. \left.
	 - 
	 \mapping_{\mathbf{w}_0} \left(\{\mathbf{h}_{\surface,\tau}, \mathbf{g}_{\surface,\tau}\}_{m\in\mathbb{M}},h_{0,\tau}\right)
	  \right\|^2_2 \right).
\end{multline}

\section{Simulation Results}\label{sec:results}
\begin{figure}[!t]
	\centering
	\includegraphics[width=0.9\linewidth]{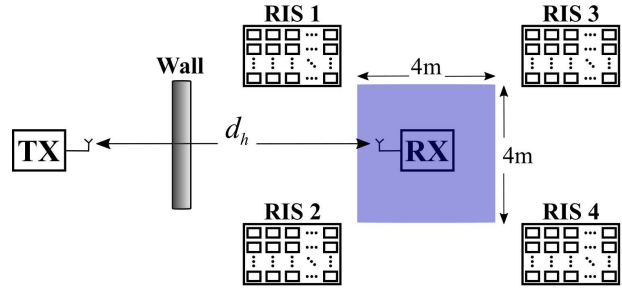}
	\caption{The general simulation setup with the placement of the TX, RX, and the four RISs as listed Table~\ref{tab:parameters}, as well as the grid RX positions used during the training phase of the proposed $3$-layer perceptron NN. When the wall is considered, a penetration loss of $10$dB is induced in the direct TX-RX link.}
	\label{fig:sims_model}
\end{figure}
\begin{table}[!t]
    \centering
    \caption{Setting of the Parameters Used in the Simulations.}
    \label{tab:parameters}
    \begin{tabular}{l c || l c} \hline
         \textbf{Parameter} & \textbf{Value} & \textbf{Node} & \textbf{Position [m]}\\
         \hline
         Number of RISs ($\SURFACE$) & 4 & TX & $(0,30,2)$\\
         Elements per RIS ($\ELEMENT$) & $8\times 8$ & RX & $(d_h,30,1)$\\
         Groups per RIS ($\ELEMENT_0$) & 4 & RIS \#1 & $(d_h-5,25,2)$ \\
         Discrete Phases ($\theta_{m,k}$) & $\{0,\pi\}$ & RIS \#2 & $(d_h-5,35,2)$\\
         TX Power ($P$) & 1W & RIS \#3 & $(x_1,y_1,2)$\\
         Noise Power ($\sigma^2$) & 100dBm & RIS \#4 & $(x_2,y_2,2)$ \\
         \hline
    \end{tabular}
\end{table}
In this section, we consider three versions of the simulation setup depicted in Fig$.$~\ref{fig:sims_model} with the parameters' setting presented in Table~\ref{tab:parameters}. In particular, we have simulated: \textit{i}) \textit{Setup 1} with $d_h=20$m, $x_1=x_2=y_1=25$m, and $y_2=35$m, where the wall is absent; \textit{ii}) \textit{Setup 2} having again no wall with $d_h=10$m, $x_1=x_2=5$m, $y_1=27.5$m, and $y_2=32.5$m; and \textit{iii}) \textit{Setup 3} which is \textit{Setup 2} with the wall inclusion resulting in a penetration loss of $10$dB in the direct TX-RX link. We have generated the following two datasets using the channel model in Section II.A: \textit{i}) a \emph{training dataset} of $5000$ channel realizations, where during each channel realization the RX was placed on a $3\times 3$ positions' square grid of width $4{\rm m}$ centered on the point $(20{\rm m},30{\rm m},1{\rm m})$ on the 3D Cartesian coordinate system; and \textit{ii}) a \emph{testing dataset} of again $5000$ channel realizations, where RX was placed randomly within the aforementioned grid for each channel realization. For comparisons of the NN-based achievable rate results with the optimum performance, we have simulated the best phase configurations for all four RISs solving \eqref{eqn:maximize_datarate} by the exhaustive search approach, which is hereinafter referred to as ``Exhaustive.'' In addition, we have simulated the performance of the baseline method ``Random,'' according to which the phase configurations of all RISs were randomly chosen without any channel knowledge or coordination. We finally term as ``No RIS'' the case where RISs are not used and only the weak direct link of length $d_h$ between TX and RX is present for their wireless communication.
\begin{table}[!t]
    \centering
    \caption{Dimensions for the NN Architecture in the Simulation Setup.}
    \label{tab:nn_params}
    \begin{tabular}{l c c c c}
    & \multicolumn{2}{c}{Position-based} & \multicolumn{2}{c}{Channel-based} \\
         \textbf{Layer} & \textbf{CEN} & \textbf{IND/FL} & \textbf{CEN} & \textbf{IND/FL} \\
         \hline
         Input & 771 & 195 & 1028 & 260\\
         Hidden \#1 & 256 & 64 & 256 & 64\\
         Hidden \#2 & 128 & 32 & 128 & 32\\
         Hidden \#3 & 16 & 4 & 16 & 4 \\
         Output & 16 & 4 & 16 & 4 \\
         \hline
    \end{tabular}
\end{table}

In the achievable rate performance results, where \eqref{eqn:datarate} was numerically evaluated, we have considered the proposed position-based NN design that is fed with the relative positions measured at each RIS element, as well as the NN design that requires as inputs the channel coefficients for all links among TX, RX, and the $M$ RISs. For the training of all simulated NNs for the parameters' setting in Table~\ref{tab:parameters}, we have used the following three approaches: \textit{i}) \emph{CENtralized (CEN)} approach that requires a central entity to realize a single large NN; \textit{ii}) \emph{INDividual (IND)} approach where each RIS implements its own NN using as inputs the measurements (positioning values or channel coefficients) referring to the channel between itself and TX as well as RX; and \textit{iii}) \emph{Federated Learning (FL)} approach: this approach extends the IND one by frequent model averaging following the federated learning concept \cite{FL_Google}. In particular, the parameters of the individual NNs during the training phase are assumed to be collected in a central entity and then averaged to be used by each individual NN in the testing phase, offering low-complexity collaborative training. The dimensions of the NN architectures (i.e., number of parameters per layer for the single large and each small individual MLPs for CEN and IND, respectively) used for all aforementioned approaches are summarized in Table~\ref{tab:nn_params}.
\begin{figure}[!t]
    \centering
    \includegraphics[width=0.97\linewidth]{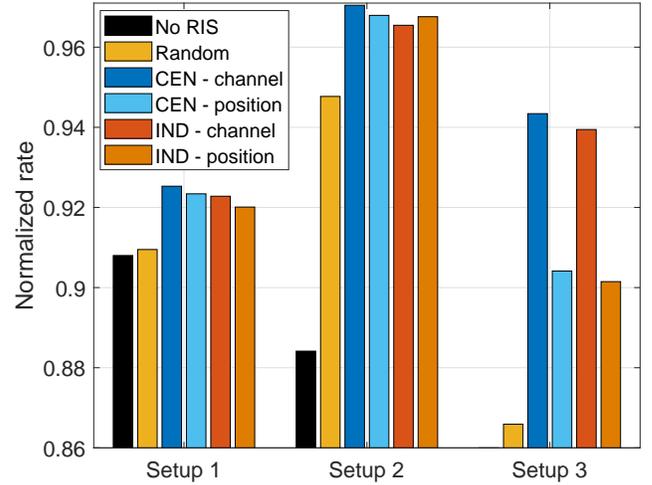}
    \caption{The normalized achievable rate performance (by the upper bound offered by the ``Exhaustive'' scheme) with all considered NN approaches for the phase configurations of all four RISs and all three distinct setups.}
    \label{fig:normRate}
\end{figure}

In Fig$.$~\ref{fig:normRate}, we illustrate the normalized achievable rate performance for the proposed NN-based approaches and the considered baseline schemes. The normalization has been calculated over the achievable rate of the ``Exhaustive'' scheme providing the best phase configurations for all RISs. We have found that the FL approaches perform very close to the IND ones, hence, they have been omitted from all figures. It can be seen from Fig$.$~\ref{fig:normRate} that all proposed NN-based designs outperform the baseline schemes ``No RIS'' and ``Random,'' with their performance improvements being dependent on the simulation setup. For example, the baseline schemes achieve around $91\%$ of the ``Exhaustive'' search rate for the Setup 1, while this value falls below $87\%$ for the Setup 3. This trend witnesses that the proposed NN-based phase configuration optimization becomes more profitable as the direct TX-RX link gets weaker. It is also evident that all proposed designs yield achievable rates that are beyond the $90\%$ of the upper-bound performance for all three investigated setups. Interestingly, both the position-based designs (i.e., CEN and IND) perform very close to their respective channel-based counterparts for Setups 1 and 2, with values around the $92\%$ and $97\%$ of the optimum achievable rates, respectively. Recall that the additional benefit of the position-based designs is that they require much lower overhead for accessing RX location compared to measuring all channel coefficients, required by the latter designs. However, it can be seen from Setup 3 that the channel-based designs are the best option, outperforming the position-based ones by $4\%$. It is also shown in the figure that the IND schemes perform very close to the CEN ones. Note that the IND schemes do not rely on frequent interactions with a central entity (with implications in control signaling overhead), as needed by both CEN schemes. 
\begin{figure}[!t]
    \centering
    \includegraphics[width=0.97\linewidth]{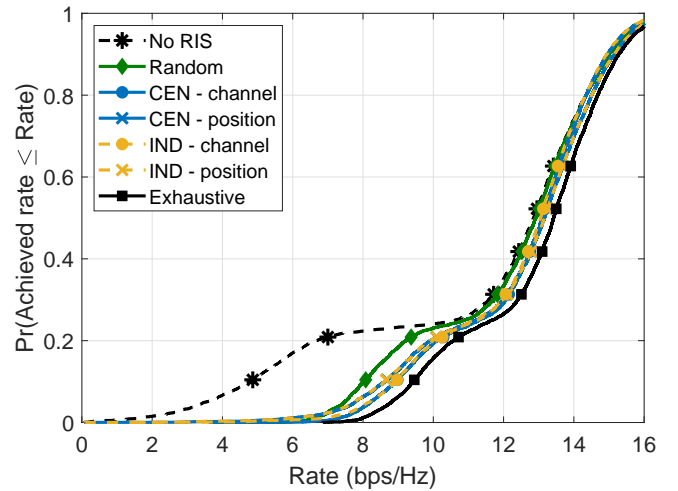}
    \caption{The outage probability of the Setup 2 (no wall between TX and RX) considering all NN approaches for the phase configurations of all four RISs.}
    \label{fig:mean_var}
\end{figure}
\begin{figure}[!t]
    \centering
    \includegraphics[width=0.97\linewidth]{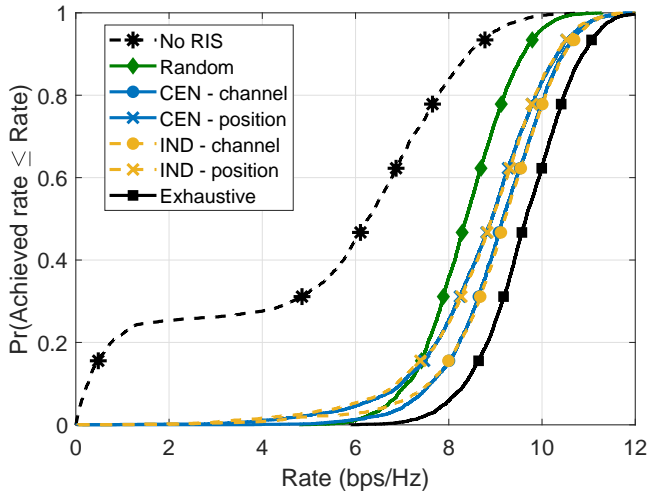}
    \caption{The outage probability of the Setup 3 (wall penetration loss of $10$dB) considering all NN approaches for the phase configurations of all four RISs.}
    \label{fig:mean_var_reduced}
\end{figure} 

The outage probability, obtained from the empirical cumulative distribution function of the achievable rate, is demonstrated in Figs$.$~\ref{fig:mean_var} and~\ref{fig:mean_var_reduced} for the Setups 2 and 3, respectively. As clearly seen in both figures, the behavior of all proposed NN-based designs is similar and sufficiently close to the optimum phase configuration. As also discussed in Fig$.$~\ref{fig:normRate}, the baseline schemes become inefficient when the direct TX-RX link gets obstructed.

\section{Conclusion}\label{sec:conclusion}
In this paper, we presented supervised learning schemes for wireless communication systems empowered by multiple passive RISs. We proposed both centralized and distributed MLP-based NN architectures, which were based either on the positioning values among the radiating elements of all involved wireless nodes or on the instantaneous coefficients of all involved wireless channels. We assessed the complexity requirements of all presented NN designs and compared their achievable rate with the optimal phase configuration based on exhaustive search. Our indicative simulation results showcased that by equipping each passive RIS with a position-based NN, achievable rates close to the optimum scheme can be achieved.     
\bibliographystyle{IEEEtran}
\bibliography{mybib_ris_letter}

\end{document}

%% file: my_definitions.tex
\newcommand{\surface}{m}
\newcommand{\SURFACE}{M}

\newcommand{\element}{k}
\newcommand{\ELEMENT}{K}

\newcommand{\action}{\phi}
\newcommand{\actionVec}{\vect{\action}}

\newcommand{\mapping}{f}
\newcommand{\modelParam}{\vect{w}}

\newcommand{\distance}{p}

\newcommand{\distanceVec}{\vect{\distance}}
\newcommand{\distanceVecIn}{\vect{\distance}^{\text{TX}}}
\newcommand{\distanceVecOut}{\vect{\distance}^{\text{RX}}}
\newcommand{\vect}{\boldsymbol}

\newcommand{\set}[1]{\mathcal{#1}}

\newcommand{\size}[1]{|\set{#1}|}

\DeclareMathOperator*{\argmin}{arg\,min}

\newcommand{\optimal}{^\star}

\newcommand{\subjectTo}{\text{subject to}}

%% file: my_acronym.tex
\newacronym{tx}{Tx}{transmitter}

\newacronym{rx}{Rx}{receiver}

\newacronym{ris}{RIS}{reconfigurable intelligent surface}

\newacronym{csi}{CSI}{channel state information}

\newacronym{mle}{MLE}{maximum likelihood estimation}

\newacronym{mmw}{mmWave}{millimeter wave}

\newacronym{pl}{PL}{path loss}

\newacronym{los}{LOS}{line-of-sight}

\newacronym{nlos}{NLOS}{non line-of-sight}

\newacronym{rss}{RSS}{received signal strength}

\newacronym{pdf}{PDF}{probability distribution function}

\newacronym{cdf}{CDF}{cummulative density function}

\newacronym{gpr}{GPR}{Gaussian process regression}

\newacronym{nn}{NN}{neural network}

\newacronym{relu}{ReLU}{rectified linear unit}

\newacronym{mlp}{MLP}{multi-layer perceptron}